%% file: main.tex
\documentclass[sigplan,10pt,screen]{acmart}
\renewcommand\footnotetextcopyrightpermission[1]{}
\usepackage{comment}
\usepackage[normalem]{ulem}
\input{header}

\begin{document}
	\title{Protecting File Activities via Deception for ARM TrustZone}
	\date{}
	\author{Liwei Guo}
	\email{lg8sp@virginia.edu}
	\affiliation{%
		\institution{University of Virginia}
		\country{USA}
	}

	\author{Kaiyang Zhao}
	\email{kaiyang2@cs.cmu.edu}
	\affiliation{%
		\institution{Carnegie Mellon University}
		\country{USA}
	}

	\author{Yiying Zhang}
	\email{yiying@ucsd.edu}
	\affiliation{%
		\institution{University of California, San Diego}
		\country{USA}
	}

	\author{Felix Xiaozhu Lin}
	\email{felixlin@virginia.edu}
	\affiliation{%
		\institution{University of Virginia}
		\country{USA}
	}

\begin{abstract}
	
	A TrustZone TEE often invokes an external \fs{}.
	While filedata can be encrypted, the revealed file activities leak  secrets.
	To hide the file activities from the \fs{} and its OS,
	we propose \sys{}, a deception-based defense injecting sybil file activities as the cover of actual file activities.
	
	\sys{} contributes three new designs.
	(1) To make the deception credible, the TEE generates sybil calls by replaying file calls from the TEE code under protection.
	(2) To make sybil activities cheap,
	the TEE requests the OS to run $K$ \fs{} images simultaneously.
	Concealing the disk, the TEE backs only one image with the actual disk while backing others by only storing their metadata.
	(3)
	To protect \fs{} image identities,
	the TEE shuffles the images frequently, preventing the OS from observing any image for long.

	\sys{} works with unmodified \fs{}s shipped with Linux.
	On a low-cost Arm SoC with \ext{} and \ffs{}, our system can concurrently run as many as 50 \fs{} images with 37\% disk overhead (less than 1\% of disk overhead per image).
	Compared to common obfuscation for hiding addresses in a flat space, \sys{} hides file activities with richer semantics. Its cost is lower by one order of magnitude while achieving the same level of probabilistic security guarantees.
	
\end{abstract}	
\maketitle

\input{introduction}

	\input{background}

\input{motiv}

\input{overview}\input{security}

	\input{design}

\input{impl}

\input{sec-analysis}

	\input{exp} %
	\input{related-work}

	\input{conclusion}

	\bibliographystyle{plain}

	\bibliography{bib/lwg,bib/xzl,bib/iot,bib/security,bib/misc}

\end{document}

%% file: header.tex
\usepackage{mathtools}

\usepackage{amssymb} %
\usepackage{wrapfig}
\usepackage{array}
\usepackage[ruled,vlined]{algorithm2e}
\usepackage{colortbl}
\usepackage{url}
\usepackage{color,soul}
\usepackage{graphics}

\usepackage{lipsum}
\usepackage{tikz}
\usepackage{enumitem}	%
\usepackage{listings} %
\usepackage{euler}

\usepackage{booktabs}	%
\usepackage{lipsum}
\usepackage{boldline}
\usepackage{amssymb} %
\usepackage{pifont} %
\usepackage{capt-of}	%
\usepackage{todonotes}
\usepackage{multirow}
\usepackage{makecell}

\usepackage{setspace}

\definecolor{labelkey}{rgb}{0,1,0}

\definecolor{airforceblue}{rgb}{0.36, 0.54, 0.66}
\definecolor{frenzyorange}{RGB}{249, 158, 26}
\newcommand*\circled[1]{\tikz[baseline=(char.base)]{
		\node[shape=circle,fill=frenzyorange,draw,text=white,inner sep=1pt] (char) {#1};}}

\renewcommand{\paragraph}[1]{\vskip 3pt\noindent\textbf{#1 }}	 %
  {\begin{list}{$\bullet$}%
     {\setlength{\parsep}{0pt}%
      \setlength{\topsep}{0pt}%
      \setlength{\itemsep}{2pt}}}%
  {\end{list}}
\newcommand\Note[1]{\sethlcolor{green} \hl{#1}} %
\newcommand\noted[1]{} %

\newenvironment{myitemize}%
  {\begin{itemize}
	[leftmargin=0cm,itemindent=.3cm,labelwidth=\itemindent,
		labelsep=0pt,
		parsep=0pt,
		topsep=1pt,
		itemsep=2pt,
		align=left]
  }%
  {\end{itemize}}    

\newenvironment{myenumerate}%
  {\begin{enumerate}
	[leftmargin=0cm,itemindent=.5cm,labelwidth=\itemindent,
		labelsep=0pt,
		parsep=1pt,
		topsep=1pt,
		itemsep=3pt,
		align=left]
  }%
  {\end{enumerate}}

  {\begin{enumerate*}
	[label=\roman*)]
  }%
  {\end{enumerate*}}

\newcommand\sect[1]{Section~\ref{sec:#1}}	%

\input{terms}

%% file: terms.tex
\newcommand{\sys}{$Enigma$}

\newcommand{\Fs}{Filesystem}
\newcommand{\fs}{filesystem}

\newcommand{\tz}{TrustZone}

\newcommand{\K}{\mbox{$K$}}

\newcommand{\ext}{EXT4}
\newcommand{\ffs}{F2FS}

%% file: introduction.tex
\section{Introduction}
\label{sec:intro}

TrustZone is the trusted execution environment (TEE) on Arm CPUs.
To use TrustZone, developers encapsulate security-sensitive code as trustlets, which are isolated in the secure world and shielded from an untrusted OS  ~\cite{trustshadow,tzslicer}.

\paragraph{File services for TEE}
Many trustlets store security-sensitive data as files, such as sensor readings and login credentials.
As shown in Figure~\ref{fig:overview}(a), trustlets often export  file calls (e.g. open/read/write/close) to the OS which hosts a modern \fs{}.
Doing so gives trustlets access to modern file features such as crash consistency and flash optimizations from various mature filesystems; meanwhile the \fs{} code does not have to be pulled into the TEE, keeping the TEE lean. %

\paragraph{Question \& challenges}
Reliance on an external \fs{} suffers from a key drawback: leak of file activities.
Although a trustlet can encrypt file contents for confidentiality and  integrity,
it has to send file \textit{activities} in the clear.
The activities include file operation types (e.g. read, write, seek, and create), sizes/offsets, and access occurrence (e.g. ``the trustlet just created a file'').
From the received file activities, the OS can infer a truslet's secrets such as input data~\cite{obliviate,sgx-lkl}.
\sect{motiv:attacks} will show evidence.

In general, access activities can be obfuscated by injecting sybil activities ~\cite{kanon,sybilquery,kloakdb}.
When it comes to file activities, existing solutions are inadequate.
(1) Popular obfuscation techniques, e.g. ORAM~\cite{oram}, focus on hiding data access \textit{addresses} in a flat, memory-like space.
While they can generate random offsets within a  file~\cite{obliviate},
they cannot generate file operations with rich semantics, %
e.g. ``read /a/index at offset 42; then open /b/data and write at offset 1024''.
(2) How to make sybil file activities \textit{credible}?
Merely making them \textit{legal}, e.g. no out-of-bound reads, is not enough to deceive an OS that has prior knowledge of the true file activities.
(3) How to minimize the cost of sybil activities,
which often amplifies the actual activities significantly?
For instance, a file backed by ORAM-like disk blocks consumes up to 10$\times$ more space and slows down access by at least one order of magnitude~\cite{obliviate}.

\sys{} is a deception mechanism that hides file activities for a TrustZone TEE.
It centers on two insights.
First, while invoking an external \fs{}, the TEE conceals the underlying physical disk\footnote{This paper uses disks to refer to storage hardware including flash.}. 
This allows the TEE to inject numerous sybil file calls but discard their disk activities covertly with little cost. 
Second, to make sybil activities credible, the TEE should borrow knowledge from the trustlet under protection.

\input{fig-overview}

\paragraph{1. Sybil \fs{}s with covert emulation}
Sybil activities are expensive as they pollute the actual data on disk.
We therefore instantiate multiple \fs{} images:
one \textit{actual} image, to which the TEE sends actual file calls; and many \textit{sybil} images, to which the TEE sends sybil calls.
This is shown in Figure~\ref{fig:overview}(b).
The separation of \fs{} images allows the TEE to fulfill their disk requests differently: performing all the disk requests from the actual image;
silently dropping \textit{filedata} accesses from the sybil images.
Essentially, the TEE emulates storage for sybil images with only their \textit{metadata},
reducing their overheads to just enough for deceiving the OS.
The TEE further implements measures against OS probing the internals of such covert emulation.

\paragraph{2. Protecting \fs{} identity via shuffling}
As the OS observes longer history of file activities, it poses an increasing threat.
For instance, the OS can determine which image may be actual by comparing the current and the past file activities on an image.
If the OS succeeds, it uncovers all the actual file activities in retrospect and in the future.

Our defense is to prevent the OS from observing file activities on any \fs{} image for long.
In the spirit of moving target defense (MTD~\cite{mtd-book}),
TEE periodically shuffles \fs{} images that have identical OS-visible states.
Not knowing the shuffling scheme, the OS can only track an image's activities for a short period of time, less than several seconds in our implementation.
TEE does shuffling efficiently by only updating metadata references, not the metadata itself or filedata.

\paragraph{3. Generating credible activities calls via replay}
The TEE should issue sybil file calls close to what the trustlet would issue; it cannot draw sybil calls, for example, from generic file traces.
The TEE can only deceive the OS when it knows the trustlet better than the OS.
The challenge is that the TEE can hardly model a trustlet's behaviors or assess how much the OS already knows about the trustlet.

Our idea is for the TEE to replay file traces pre-recorded from the very trustlet to be protected.
The file traces hence form a tight envelope of the trustlet's actual file activities.
\sys{} provides support for developers to collect file call segments and for the TEE to produce an unbounded stream of sybil calls at run time.

\paragraph{Results}
By constraining lightweight modifications to generic subsystems and interfaces, Enigma eschews heavy internal changes to individual filesystems and works with unmodified EXT4 and F2FS, reusing over 60K SLOC filesystem-specific implementations.
Through a study of six diverse trustlets from which we collect over 200K file calls through testing, we show \sys{} is practical to deploy and effectively hides the file activities that leak trustlet secrets.

\sys{} provides the following guarantees: 
(1) 
Against random guess attacks: 
the probability of a successful guess is 1/K; 
the successes of individual attacks are independent. 
(2) Against an external, persistent observer: 
the maximum period of continuously observing any \fs{} image is T. 
Both K and T are user-configurable. 

On a low-cost ARM board (RaspberryPi 3) running 20 concurrent \fs{} images,
\sys{} incurs 2.2$\times$ access slowdown and consumes 25\% additional disk space (with 1.5 MB per sybil image on average);
with as many as 50 concurrent \fs{} images,
\sys{} incurs 3.9$\times$ access slowdown and 37\% additional disk space.

\paragraph{Contributions}
This paper presents \sys{}, a novel mechanism that generates credible, rich sybil file activities at low cost.
\sys{} contributes the following new designs:

\begin{myitemize}
\item
Sybil \fs{} images emulated with only their metadata,
which makes strong deception with numerous sybil file activities affordable.

\item
Continuous shuffling of \fs{} identities, which prevents an external observer from collecting long histories of file activities, sybil or actual.

\item
Replaying file call segments recorded from the trustlet under protection, which effectively deceives a knowledgeable observer.

\end{myitemize}

For a TrustZone TEE, \sys{}'s deception approach opens the door to using more untrusted external OS services. %

%% file: fig-overview.tex
\begin{figure}[t!]
	\centering
	\includegraphics[width=0.48\textwidth{}]{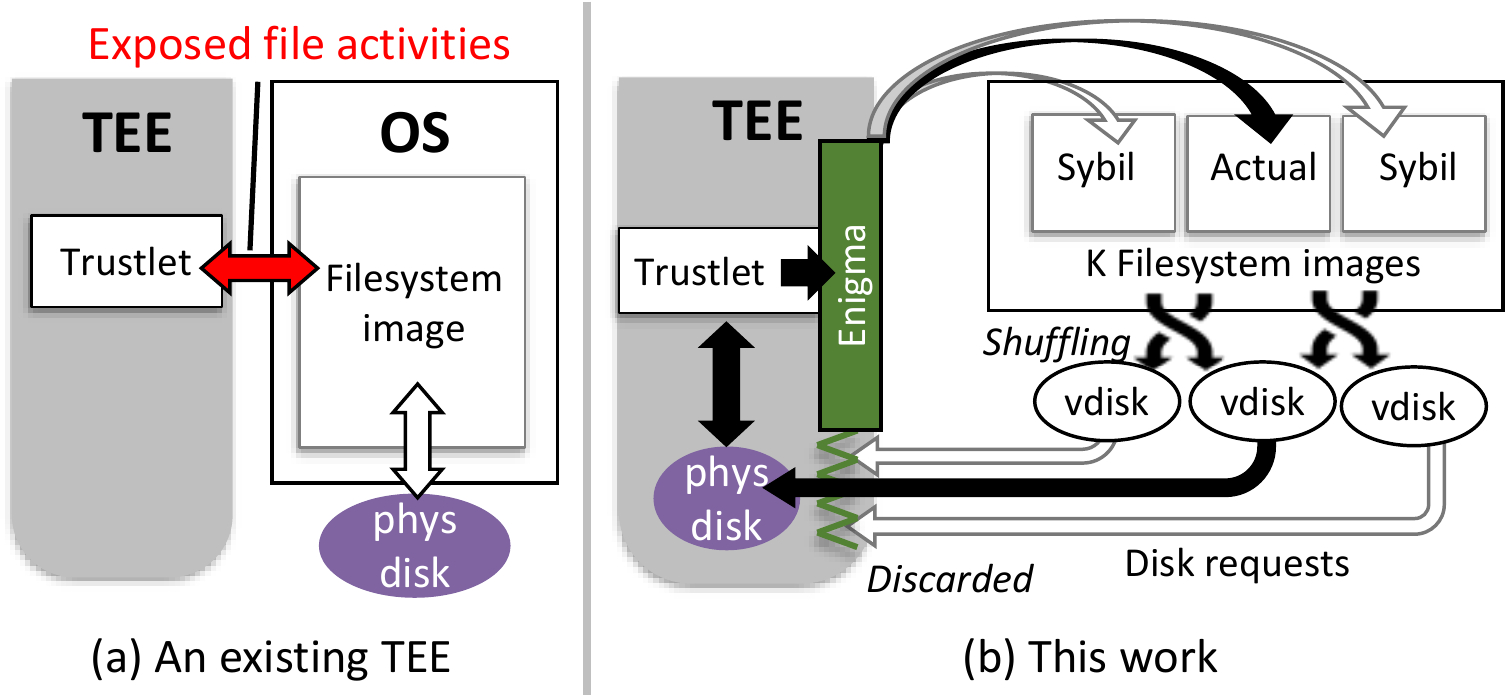}
	\caption{An overview of this work}
	\label{fig:overview}
\end{figure}

%% file: background.tex
\section{Motivations}
\label{sec:background}
\label{sec:bkgnd}
\label{sec:motiv}

\subsection{\tz{} and its file services}
\label{sec:tee}
\paragraph{TEE secure storage}
Arm \tz{} statically partitions an SoC's physical memory and IO devices between normal world and a secure world (i.e. TEE)~\cite{santos}. 
The TEE can isolate storage medium, e.g. a SD card or a flash partition, from the normal world OS. 
The resource partitioning differs from Intel SGX where memory is mapped to TEE dynamically and the OS controls IO hardware. 

\paragraph{TEE needs file services}
Mobile/embedded devices produce and store security-sensitive data such as 
user health logs and audios samples. 
TEE can isolate sensitive data from high-risk software such as the OS, for which TEE
needs a modern \fs{} to keep the data persistent.
For instance, journaling~\cite{journaling-fs} prevents data corruption, which is not uncommon on battery-powered smart devices; 
wear-leveling extends flash lifespan~\cite{wear-leveling-survey}, which is key to flash longevity as IoT devices or their flash can be difficult to replace. 

Unfortunately, modern \fs{}s are complex, making them unsuitable to run within the TEE. 
First, a modern \fs{} has substantial code. 
For instance, EXT3/4 and F2FS have 35K and 17K SLOC respectively. 
They would significantly bloat TEE's trusted computing base (TCB), 
e.g. the popular OPTEE which only has around 25K SLOC.
Filesystem vulnerabilities~\cite{fs_evo,kernel_bug} then become attack vectors against the TEE. 
Second,
lifting-and-shifting a modern \fs{} to TEE requires tedious effort.
Not only the \fs{} but also extensive kernel APIs must be ported, e.g. VFS, page allocation, and workqueues. 
Trimming the \fs{} code for TEE is error-prone, for which developers need to  thoroughly understand \fs{} logic and test rigorously. 
Maintaining a \fs{}'s separate versions for OS and TEE complicates distributing security updates and patches, which may give rise to a fragmented ecosystem.

For these reasons, forwarding file calls to the OS is a common practice of trustlets.

\subsection{The Linux storage stack}

Our design exploits the following OS storage features. 

\paragraph{The stack layers}
At the top of the stack, the virtual \fs{} (VFS) is a \fs{}-agnostic frontend receiving \fs{} calls, such as read or write, from \fs{} clients. 
VFS dispatches \fs{} calls to concrete \fs{} implementations; VFS also caches recent file access. 
A \fs{} translates the file calls to disk requests, e.g. block read/write, and submits the requests to an underlying block layer.

\paragraph{Filedata vs. Metadata}
A \fs{}'s all on-disk state constitutes its \textit{image}. 
The image consists of filedata as user contents and metadata which describes the file and the \fs{}. 
Common metadata examples are inodes, directory structures, and block maps. 
Metadata often constitutes a small fraction of \fs{} image, a premise to be tested in \sect{eval}.
To execute a file call, a \fs{} often examines the metadata, e.g. 
reading inodes of a file in order to locate the disk blocks.

\input{threat-model}

%% file: threat-model.tex
\subsection{The attacks}
\label{sec:motiv:attacks}

\paragraph{Threat model}
We trust the software in TEE. 
Both the TEE's file contents and file activities may expose its secret. 
The file activities are driven by TEE software only and \textit{not} by untrusted entities, e.g. a normal-world app communicating with the TEE. 
The filedata and file/directory names can be encrypted by the TEE.

The OS hosts a \fs{} for the TEE. The OS is:
\begin{myitemize}
\item

\textit{Curious.}
The OS probes the TEE's secrets passively and actively. 
(1) It monitors the TEE's file activities, including file calls, disk requests, data move, and timing of these activities; 
(2) it inspects a \fs{}'s in-kernel state; 
(3) it may deviate the \fs{} logic to request disk reads or writes. %

\item 
\textit{Knowledgeable.}
The OS knows the sequences of file calls that the trustlet may issue.  

\item 
\textit{With unbounded memory.}
The OS can memoize the full histories of file calls and disk requests it has ever observed. 
Following the TrustZone convention~\cite{tz-hw-sec}, we deem hardware attacks (e.g. bus snooping) and their side channels out of scope.

\end{myitemize}

%% file: motiv.tex
\paragraph{Side channels through file activities}
A file call exposes the following information that cannot be easily obfuscated: 
file call types~\cite{posix-guide}; 
accessed file paths, in particular the relative location in a directory tree; 
arguments, e.g. sizes, offsets, and flags.
A trace of file calls is known to leak the caller's secret~\cite{obliviate,controlledChannel,access-patten-privacy}.  

We identify three common side channels from file traces: 
 (1) \textit{Occurrence}: the events that a trustlet access files; 
 (2) \textit{File paths}; %
 (3) \textit{Access patterns}: the combination of access offsets, sizes, and flags in a sequence of file calls.

\paragraph{Trustlets \& attacks}
We motivate our designs with the following trustlets, including their side channels and secrets. Table~\ref{tab:benchmarks} shows a detailed summary.

\begin{myitemize}
\item \textit{Databases} for embedded environments such as SQLite manage on-device user data~\cite{android-fs-usage}.
Prior work shows a database's file access patterns depend on queries~\cite{obliviate,access-patten-privacy,oblidb}. 
For instance, given a database's schema and a sequence of file offsets, 
the OS can learn a query's secret: read 64 bytes at offset 0 and read 128 bytes at offset 4096 gives away page-align predicate columns (e.g. ``user physical activities'') and the rows selected (e.g. ``hours when the user is sleeping'').

\item \textit{Fulltext search} is for on-device QA over emails or messages~\cite{deqa}. 
Given a keyword, an engine reads a binary index file, locates file offsets where the keyword appears, and reads in contextual lines. 
From the access offsets, 
the OS can infer the secret: the keyword and the hit locations ~\cite{access-patten-privacy}.

\item 
\textit{Model loading.}
A trustlet loads a neural network model from a file, for which
it may issue thousands of file calls to read and parse model layers. 
As reported by prior work~\cite{nn-access-pattern} and verified by us on TensorflowLite~\cite{tensorflow-web}, 
from the file path and offsets the OS can learn the secret: the loaded model.

\item \textit{Video surveillance}. 
A trustlet on a camera detects video events of interest, e.g. motion or vehicles, and saves video frames of interest. %
The OS observes file writes and learns the secret: occurrences of events being detected. 

\item \textit{Data historian}. 
A trustlet on a robot
collects sensor messages to a ROSBag file~\cite{rosbag}.
The messages are of variable lengths depending on sensor types 
(e.g. point clouds and sound samples) as well as data contents (e.g. a point cloud's density). 
From the write sizes, the OS can learn the secret: sensor types, data contents, and hence the robot's activities.

\item \textit{Credential manager}. 
A trustlet loads one of multiple key files for authentication with a remote server. 
From the file path, the OS can infer the secret: the loaded key, which corresponds to user identity (e.g. a specific private key) or server identity (e.g. a specific CA certificate).

\end{myitemize}

%% file: security.tex
\subsection{System Overview}

\label{sec:overview}
\input{fig-arch}

Figure~\ref{fig:arch} shows the system architecture. 
\paragraph{Initialization}
With secure IO, the TEE isolates the physical disk and exposes $K$ virtual disks to the normal world. 
From the OS's perspective, a virtual disk is no different than a physical block device, except that the OS submits block requests to the TEE.
The TEE requests the OS to initialize $K$ images with the same \fs{} implementation on the $K$ virtual disks. 

\paragraph{Operation}
As a trustlet emits a stream of actual file calls, 
the TEE generates additional K-1 streams of sybil calls \circled{1} by 
replaying pre-recorded file call segments (\S\ref{sec:sybil}).
The TEE sends the actual file calls to the actual \fs{} image and the sybil calls to their respective sybil images. 

The OS runs K \fs{} images with an unmodified \fs{} implementation (although the kernel's generic storage subsystem is lightly modified, see \sect{impl}).
At the end of each file call execution, the OS submits disk read/write requests to their corresponding virtual disks in the TEE \circled{2}. 
The TEE omits all \textit{filedata} accesses from the sybil \fs{} images 
and only executes their \textit{metadata} accesses \circled{3}, which is needed by the \fs{} logic. 
The TEE periodically unmounts \fs{} images, shuffles the virtual disks backing them, and remounts the images on the shuffled virtual disks.

To probe the virtual disk internals, the OS will attempt to: 
(1) read filedata. 
(2) tamper with filedata. 
(3) measure delays of disk access.
The TEE implements mechanisms to block these attempts. 

\paragraph{What the OS can and cannot see?}
The OS sees K virtual disks exported by the TEE, on which K \fs{} images are mounted. 
It does not know which image is actual. 

The OS sees K streams of file calls sent to the K images. 
In each stream, the OS can see all file calls in the clear, including their types and arguments. 
The OS cannot access filedata referenced in file calls. 
The patterns of file calls fit in the OS's prior knowledge about the trustlet's file activities. 
All streams show similar statistics, including throughputs of file calls and bytes read and written. 

To execute file calls, the OS can freely access metadata (e.g. inode) on virtual disks but not filedata, as such attempts are blocked by TEE; 
the delays in accessing metadata are the same across all virtual disks. 

From time to time, 
the OS sees: the TEE unmounts some images and takes some virtual disks offline; 
the TEE puts online new virtual disks with random names. 
The OS cannot associate any new virtual disks to those disappeared. 

\paragraph{Limitations}
To simplify security reasoning,
we consider that one actual \fs{} image is exclusively used by one trustlet;
concurrent trustlets use separate \fs{} images.

%% file: fig-arch.tex
\begin{figure}[t]
	\centering
	\includegraphics[width=0.45\textwidth{}]{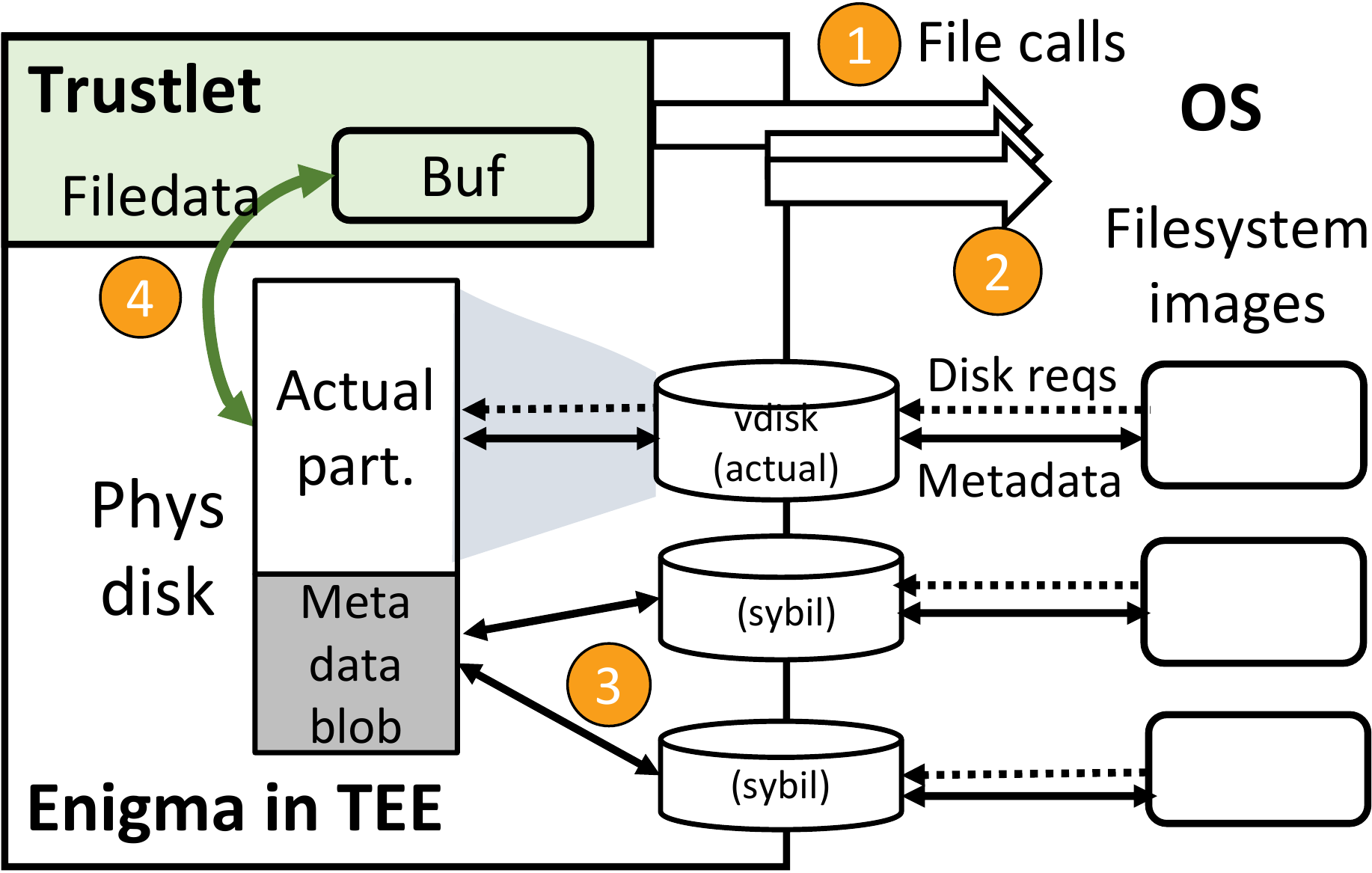}
	\caption{The system architecture, showing the filedata/metadata separation and the physical disk layout}
	\label{fig:arch}
\end{figure}

%% file: design.tex
\input{design-disk-xzl}

\input{design-disk-migrate-lwg}

\input{obfuscator-op-generation}

%% file: design-disk-xzl.tex
\section{Sybil images with covert emulation}
\label{sec:disk}

We seek to minimize the cost of sybil images so that \sys{} can afford a lot of them for strong protection.

\subsection{Metadata-only sybil images}
\label{sec:disk:path}

The first question is what are the minimum disk requests for maintaining a sybil image?
We exploit a \fs{} invariant: 
a \fs{} only relies on its \textit{metadata} to function properly (e.g. to read or update inodes)~\cite{david}. 
Note that the filedata/metadata division in a modern \fs{} can sometimes be ambiguous, for which we will describe treatment (\S\ref{sec:impl}). 
\sys{} therefore enforces that the code of \fs{} and OS only access metadata content and its access status (e.g. write completion), but not those of the filedata. 

The second question is how to hide the fact that TEE only stores metadata for sybil images?
The TEE conceals the physical disk from the OS using the TrustZone's secure IO (Section~\ref{sec:bkgnd}). 
Note that the OS still manages insecure storage devices out of TEE. 
As shown in Figure~\ref{fig:arch}, 
the TEE directly backs the actual image with a contiguous physical disk region. 
This preserves the locality of actual disk accesses. 
The TEE stores the metadata for all sybil images in a single binary blob.
To save disk space, it compresses the metadata with copy-on-write (CoW).
CoW is effective because the sybil images are mutated by file calls from the same trustlet and are likely to share similar metadata.

We next describe the OS's probing attacks against the covert emulation, as well as our defense.

\subsection{Isolating filedata paths}
\label{sec:disk:isolate}
\textbf{Threat:} 
Knowing sybil \fs{}s contain only metadata, the OS submits disk read requests for filedata on virtual disks; 
it knows the disk as sybil if no data comes out.

\noindent
\textbf{Defense:}
TEE mitigates the threat by isolating the filedata path, which keeps the OS oblivious to \textit{if} filedata disk requests are actually executed on disk, or \textit{when} such requests are completed.
The isolation is as follows.

\noindent
{1. The filedata path:  trustlet $\leftrightarrow$ disk.} \hspace{1mm}
Filedata flows between a trustlet and the physical disk without leaving the TEE. 
As shown in Figure~\ref{fig:arch}, 
a trustlet makes file calls with opaque references of filedata buffers (e.g. ``read from offset 10 to buffer $<$a 64-bit int$>$'') (\circled{1}); 
the OS executes the file calls and issues disk requests containing these opaque references (\circled{2}); 
the TEE receives the disk ops, maps the opaque references back to the filedata buffers, and moves the filedata between the trustlet and the disk (\circled{4})  without going through the OS. 
To prevent fabrication, each opaque reference is a 64-bit integer and used one time only~\cite{opaque,streambox-tz}.
	
\noindent	
{2. The metadata path: OS $\leftrightarrow$ disk.}  \hspace{1mm}
Metadata flows between the OS and the TEE's disk. 
As shown in Figure~\ref{fig:arch} (\circled{3}): 
the \fs{} code generates disk requests for metadata (e.g. ``write to an inode at block 42''), 
which contains cleartext references to OS buffers.
The TEE executes the requests and copies metadata between the TEE and the supplied OS buffers.
The TEE notifies the OS of metadata access completion. 
The TEE never examines metadata or takes any action based on the metadata content.

\subsection{Rejecting OS access to filedata}
\label{sec:disk:reject}

\textbf{Threat:}
The filesystem code, by design, may rightfully repurpose disk blocks without notifying the underlying disk.
For instance, after flushing the redo log (metadata), \ext{} may store filedata to the underlying blocks~\cite{ext4-forensic}. 
A malicious OS may tamper with the repurposed filedata blocks, breaking filedata integrity and/or revealing the identity of sybil images (i.e. those containing no proper filedata after filedata writes).

\noindent
\textbf{Defense:}
The TEE rejects OS from accessing filedata.
To do so, it tracks filedata blocks on the physical disk and keeps the filedata/metadata dichotomy up to date.

Without requiring intimate \fs{} knowledge, TEE enforces a simple policy: 
\textit{allow the OS to only read back disk blocks it previously wrote to}. 
When a block is written initially, the TEE tags the block as ``metadata'' or ``filedata'' depending on whether the written data 
comes from an OS buffer or a trustlet buffer. 
The TEE accommodates block repurposing: 
in case data is written from an OS buffer to a ``filedata'' block, 
the TEE erases the block so no existing content is leaked, changes its tag as ``metadata'', and grants the access. 

Some \fs{}s may inline filedata in metadata blocks for efficiency, e.g. \ext{}  may inline files smaller than 160B. 
The TEE tags these blocks as ``mixed'' and tracks inlined filedata ranges. 
As a result, the TEE allocates disk space for inlined filedata on sybil images, which incur less than 1\% of the space overhead in our measurement.

\subsection{Defense against timing attacks}
\label{sec:disk:timing}

\noindent
\textbf{Threat:}
The OS measures disk delays of metadata access.
Since a sybil image's metadata is stored more compacted than the actual image, the OS may see lower access delays. %

\noindent
\textbf{Defense:}
The TEE pads delays of metadata access so that the OS sees uniform delays. 
Note that it does not have to pad delays for filedata access, 
the completion notifications for which bypass the OS as shown in Figure~\ref{fig:arch}.

The TEE delays each metadata access to all \fs{}s to be longer than most (e.g. 99\%) of the actual delays, a value determined by profiling when \fs{} is mounted.
Delaying is practical for two reasons.
i) Metadata accesses only constitute a small fraction of all disk access. 
ii) The actual access delays on embedded flash show low variation because the storage has limited internal buffering. 
We will show timing side channel reduction and overhead in evaluation.

%% file: design-disk-migrate-lwg.tex
\section{Filesystem identity shuffling (FIDS)}
\label{sec:fsnr}

TEE obfuscates file call histories in the spirit of moving target defense~\cite{mtd-book}. 
This is because 
an OS collecting long histories of file calls poses two threats.
(1) 
The OS is more capable of attacks. 
It can uncover \fs{} identities by reasoning about the histories. 
(2) 
The OS, if accidentally discerns, the actual \fs{}, creates higher damage. 
It learns all the actual file calls in the past and in the future.

\subsection{The mechanism}

Critically, FIDS follows an \textit{egalitarian} principle.
The TEE frequently shuffles identities of all \fs{}s, sybil and actual.
A \fs{} image's OS-visible state is $\langle v,M \rangle$: 
$v$ is the name of virtual disk that the image is mounted on; 
$M$ is the metadata, 
which is exposed to the OS by design. 

\textit{Shuffling.} \hspace{1mm}
Shuffling prevents the OS from connecting segments of file call histories. 
It is performed on a set \fs{} images 
$\{\langle v_1,M \rangle , ... , \langle v_n,M \rangle \}$
that currently have identical metadata $M$, where $n$ is the number of \fs{}s.
TEE assigns each backing virtual disk a new, random name: 
$\{\langle v'_1,M \rangle , ... , \langle v'_n,M \rangle \}$.
Since the OS cannot connect new disk names $v'_i$ to the old names $v_i$,
it cannot connect new \fs{} identities to the old ones. 

TEE triggers shuffling by time 
(e.g. an image has not participated in shuffling for a period of $T$) 
or by activities (e.g. the image has served $N$ file calls since its last shuffling).
\sect{eval} will evaluate the impact of $T$, 
and show that even in the worst case (i.e. accidental filesystem identity exposure), a practical $T$ (e.g. a few seconds) leaks no significant secrets of the trustlet.

\textit{Forking.} \hspace{1mm}
Forking keeps shuffling going when all images have distinct metadata.
In case TEE attempts to shuffle an image  $\langle v,M \rangle$ but no other images have metadata $M$, TEE creates an image with metadata $M$ and shuffles the two images, resulting in $\langle v',M \rangle$ and $\langle v'',M \rangle$.
The OS cannot connect either of the two new identities to the old image $\langle v,M \rangle$.

\textit{Retiring.} \hspace{1mm}
TEE deletes an image to reclaim its blocks.

The mechanism is inexpensive. 
First, shuffling and forking only manipulate \textit{references} to metadata, not metadata itself nor filedata. 
Section~\ref{sec:impl} presents more details.  
Second, TEE does not need to execute forking often, as 
there often exist abundant images with the same metadata. 
For instance, a read-most workload only mutates metadata occasionally.

FIDS operations are visible to the OS, e.g. after mounting, the OS knows images with identical metadata may have been forked. 
FIDS, however, does not leak filesystem identities, because both actual and sybil images can be forked and shuffled; although the actual image cannot be retired, \sys{} ensures that such a behavior does not leak identity, as will be discussed in \S\ref{sec:fsnr:extinct}.

\subsection{Why FIDS works}

We use \textit{lineage} to describe an image's OS-visible history. 
A lineage starts with one of the initial \K{} images. 
It includes descendant images created by forking and \textit{OS-perceived} descendants created by shuffling. 
Figure~\ref{fig:fsnr} shows an example: 
the first forking on image B creates two descendant C and D; the subsequent shuffling of C and D result in images E and F. 
C--F all belong to the lineage of B.
After shuffling A and F, the resultant G and H belong to both the lineages of A and B.

\paragraph{FIDS limits continuous observation} 
The OS can track the history of any image (e.g. by tagging them with unique metadata), but only for a continuous period no longer than $T$. 
This prevents the OS from collecting large number of samples (e.g. a few thousand~\cite{template-attack}) and building statistical models for filesystem images, thwarting template attacks.
Consider the actual image B in Figure~\ref{fig:fsnr}. 
Without FIDS, B has a straight-line lineage and its full history of file activities are exposed to the OS. 
While FIDS does not change the true history of B (annotated with $\bigstar$) which is only known to TEE, it makes the history appear uncertain to the OS. 
Forking adds branches to B's lineage;  
shuffling merges and then splits lineages. 
OS can only see a lineage tree that clouds over the actual history.

\paragraph{FIDS confines damage of identity exposure}
Assume the OS, via an unexpected channel, discovers image B as actual. %
Without FIDS, the OS can track backward in time and reveal all the past file calls on B; it can track forward to learn all future calls on B. 
All the actual file calls are hence leaked. 
FIDS prevents the OS from backtracking no further than the most recent shuffling event; 
earlier than that event, all the images participating in shuffling become the probable ancestors of B. %
Similarly, the OS can forward track no further than the next shuffling/forking event on B. 
\sect{eval} will quantify the resultant uncertainty to the OS.

\subsection{Defense against extinct lineage attacks}
\label{sec:fsnr:extinct}

\input{fig-fnsr}

\paragraph{Threat:} 
While forking and shuffling are egalitarian, retiring cannot be: 
TEE never retires the actual image, 
giving the OS a chance to weaken or break the actual image's anonymity. 
If the descendants of an earlier image X have all retired (i.e. an \textit{extinct} lineage), 
the OS can rule out X from being actual. %

Figure~\ref{fig:fsnr} shows an example: if the TEE retires image A prior to its renaming with F, the only remaining lineage is the one from B, which the OS can deem as the actual. 
A successful OS pinpointing the actual image thus exposes the \fs{} identity as described above.

\paragraph{Defense:}
TEE picks images to retire by respecting two invariants: 
(1) at any moment in history, there are always $K$ alive \textit{lineages} stemmed from the initial $K$ images; 
(2) the retiring event leaves at least $K$ images alive. 
The first invariant ensures that the OS's backtracking cannot rule out any of the initial image from being actual. %
The second invariant ensures the strength of anonymity at any time in history. 
Both invariants combined, the OS cannot find a time in the past when there were fewer than $K$ lineages.

To enforce the two invariants, a challenge is to avoid memoizing all FIDS events which grow unbounded. 
The TEE implements a simple rule: 
avoid retiring an image if it is the last surviving image on a lineage. 
To do so, the TEE only keeps K tags: 
tagging each image with its ancestor as one of the K initial images and propagating the tag to descendants. %

The above design will not result in too many images of which none can be retired. 
As long as there are more than K images, there are multiple images belonging to the same lineage; retiring any will satisfy both invariants above.

%% file: fig-fnsr.tex
\begin{figure}
\centering
	\includegraphics[width=0.48\textwidth{}]{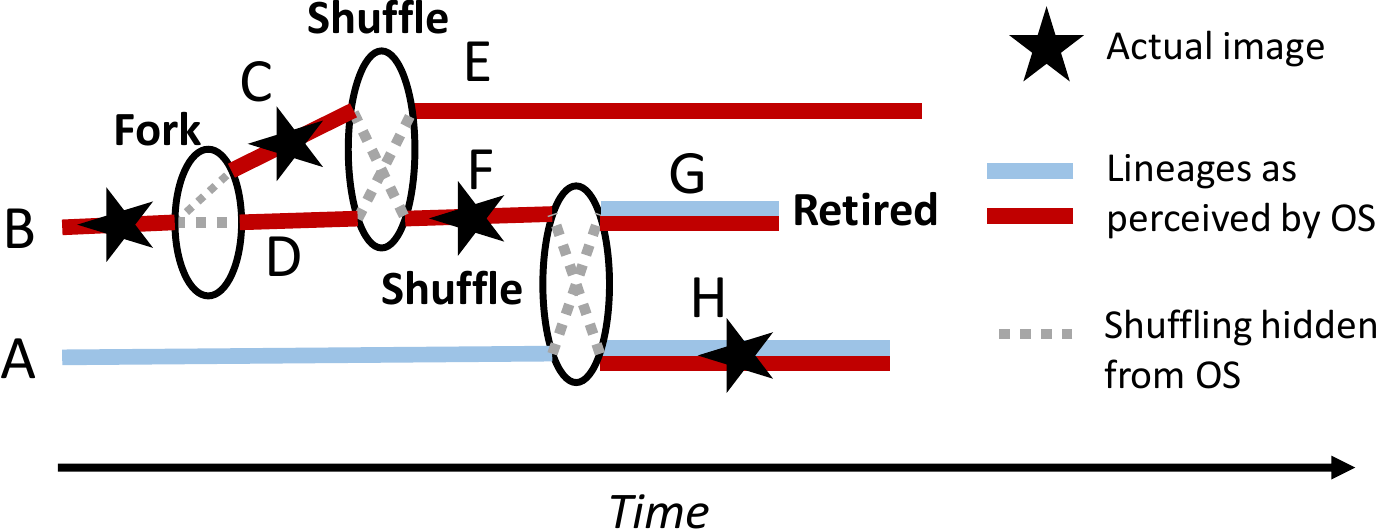}
	\vspace{-2em}
	\caption{A minimal example of \fs{} identity shuffling.
	As time goes by, the OS perceives multiple lineages as actual but cannot distinguish them
	}
	\label{fig:fsnr}
	\vspace{-1em}		%
\end{figure}

%% file: obfuscator-op-generation.tex
\section{Generating sybil file calls}
\label{sec:sybil}

We ensure that TEE generates sybil file calls close to what the trustlet would actually issue, from which the OS is unable to discern the actual file calls.
The objective is \textit{not} to match the actual file traces in deployment, but to generate diverse trace segments that provide strong cover. 

\subsection{Design}

\textbf{First, how to fit sybil file calls in a trustlet's envelope of file activities}? %
For instance, a database may show a variety of file access patterns depending on queries;
one pattern can be ``read 8 bytes, skip 16 bytes, and read 42 more bytes''. 
If a stream of file calls do not show any such pattern that the OS knows must exist in actual file calls, the OS can determine the stream of file calls as sybil. 
However, it is difficult for \sys{} to model a trustlet's file activities 
or assess how much the OS knows about the activities.

Our solution is to exploit the knowledge already encoded in a trustlet:
the TEE replays historic file traces from the trustlet to be protected. 
To this end, developers exercise the trustlet with a set of inputs and record file traces during the execution, which we will show in \sect{trace:case}.

\textbf{Second, how to generate sybil calls that provide \\strong protection?}
Our insight is that the efficacy of sybil calls hinges on \textit{the set of plausible secrets} they represent as cover traffic for the \textit{actual} secret;
we measure such efficacy as the set's cardinality and entropy estimation~\cite{beirlant1997nonparametric}.
Intuitively, a library of sybil calls would offer stronger anonymity if the library represents more plausible secrets and these secrets are uniformly distributed in the space of secrets. 

To quantify the set of plausible secrets, we exploit an observation: 
a trustlet's file calls are driven by input events~\cite{android-fs-usage,bluefs}, which are associated with the trustlet's secret. 
Therefore, we retrofit the idea of viewing a file trace as independent segments, where each trace segment encodes a secret value. 
For instance, the file trace of a database can be segmented per query and each segment encodes a secret $\langle C,R \rangle$: a query's predicate columns ($C$) and its selected rows ($R$).

With the above rationale,
\sys{} assists developers to collect sybil trace segments. 
It requires the developers to (1) provide annotations for segmenting file traces, e.g. by input events; (2) provide test inputs, such as concrete database queries; (3) annotate the inputs with plausible secrets they represent. 
A test harness exercises the trustlet and records the resultant trace segments. %
It reports cardinality and entropy of the current secret set~\cite{beirlant1997nonparametric} and makes suggestion towards improving them. 
The developers finish collection when they are satisfied with the metrics.

For example, to collect trace from a database trustlet, 
the developers provide as input a set of queries, each annotated with a secret $\langle C,R \rangle$ for the query. 
The set of plausible secrets is therefore 
$\{ \langle C_1,R_1 \rangle, \langle C_2,R_2 \rangle, ... \}$. 
After running a batch of queries and collecting the trace segments, the test harness suggests to increase the secret set entropy by 
running more queries that select more diverse columns.

\textbf{Third, how to replay the segments?} 
TEE replays by sampling from a library of trace segments;
it preserves both the pre-recorded order and arguments of file calls within a segment.
As it runs, it gradually renews the pre-recorded segments with segments (both sybil and actual) collected in deployment. 
As a result, the sybil traces evolve to be even closer to what the trustlet is issuing in deployment. 
We further address the following issues.
\textit{(1) Time the emission of segments. }
TEE emits at random intervals so that sybil file call throughput and read/write throughputs approximate that of the actual trace. 
\textit{(2) Delay between file calls. }
To prevent timing side channel, 
the TEE uniformly pads all the intervals between file calls to the maximum interval it has observed. 
\textit{(3) Make sybil calls consistent with \fs{} images. }
The TEE adjusts sybil calls before replay, 
for instance, to create files, to truncate the offsets of out-of-bound access, to redirect access from non-existing files.

\subsection{Case study}
\label{sec:trace:case}

We study the trustlets in Table~\ref{tab:benchmarks}. 
Our input for recording should be seen as examples; 
developers are likely to have inputs better matching their deployment, e.g. queries from their deployed databases or logs from their robots. 

\noindent
\textit{Database}. \hspace{1mm}
We run SQLite on a database of user health activities with 3 columns in numeric types. 
We run a suite of queries~\cite{sqlite-speedtest} and segment file traces by query. 
We collected 500 trace segments constituted by 15K file calls. 

\noindent
\textit{FullText.} \hspace{1mm}
We run Lucy, an embedded search engine~\cite{apache-lucy} over 2GB of emails~\cite{enron}. 
The inputs are 100 searches for top keywords. 
We segment file calls by search. 
The collected 100 segments consist of 100K file calls.

\noindent
\textit{ModelLoad.} \hspace{1mm}
We run TensorflowLite. 
Our inputs are 10 sample NN models loaded for inference. 
We segment file traces by each model load. 
We collected 100K file calls in 10 segments. 

\noindent
\textit{VideoEvent.} \hspace{1mm}
We run an OpenCV motion detector. 
Our inputs are 100 hours of street camera videos in Bangor~\cite{bangor-webcam}. 
We segment file calls by per video hour and have collected 9K file calls in 100 segments. 

\noindent
\textit{Historian.} \hspace{1mm}
We run the ROSBag drive data historian with 10 different drives from the autonomous driving dataset~\cite{rosbag}. 
We instrument the run script to segment file calls by per test drive. 
We have collected 9K file calls in 9K segments. 

\noindent
\textit{CredLoader} \hspace{1mm}
Our test script generates 50 key files and invokes the Openssh client on these key files for login.
We instrument the test script to segment file calls by each login attempt.
We have collected 300 file calls in 100 segments.

%% file: impl.tex
\section{Implementation}
\label{sec:impl}

We implement \sys{} in 2K SLOC, atop OPTEE and Linux as summarized in Table~\ref{tab:plat}.
Of the code, 1K SLOC is for modifying the generic kernel page cache and block subsystem; filesystem-specific code incurs only less than 50 SLOC of changes.
In another 1K SLOC, we use the MMC driverlet inside TEE~\cite{driverlet}, 
which provides read/write functions sufficient to our needs. 
We use a 32-GB microSD for storage and partition it into two: 
one is 4GB and used by the untrusted OS as its rootfs;
the other is managed by \sys{} as the isolated physical disk.
We next describe implementation details of Enigma -- how we apply lightweight instrumentations to generic kernel subsystems and avoid heavy modifications to individual filesystems.

\paragraph{TEE $\leftrightarrow$ OS interfaces}
We instrument two interfaces for communicating between TEE and OS. 
At them, we inject SMC instructions and handle world switches.
\begin{myenumerate}
	\item 
	TEE $\rightarrow$ OS. 
	Via the interface, \sys{} issues file calls (actual and sybil) to OS. 
	To this end, we instrument filesystem syscalls (e.g. generic\_perform\_read) at VFS layer. 
	We modify the data buffer pointer of the interface (i.e. iov\_iter) to pass opaque references pointing to in-TEE buffer addresses (\S\ref{sec:disk:isolate}) instead of userspace addresses;
	we further preserve them in the kernel page struct, which we will describe shortly. %
	
	\item 
	OS $\rightarrow$ TEE.
	At the \fs{} bottom, we instrumented the block IO (bio) interface (e.g. submit\_bio).
	It dequeues bios to TEE; 
	when it does so, we retrieve the opaque references from the page struct pointed by the bio and pass them back to TEE.  
	We also modify bio callbacks to let the \fs{} execute asynchronously w.r.t. TEE invocations and disable bio merging.
\end{myenumerate}

\paragraph{Isolating filedata path}
While being conceptually independent of each other, \fs{}s work closely with kernel memory management (MM) layer.
For instance, a \fs{} is also responsible for reading filedata into the page cache, and in coordination with the MM layer, writes dirty pages (filedata) back to the disk.

To isolate the filedata path described in \sect{disk:isolate}, we disengage the page cache layer as follows.
We first modify the page cache allocation methods (e.g. pagecache\_get\_page) to preserve the opaque references in page cache. 
By design, OS must allocate page cache before manipulating data pointed by foreign addresses (e.g. by userspace or opaque references).
At TEE $\rightarrow$ OS interface when OS allocates pages, we tag all newly allocated pages and inherit opaque references in their page struct.
We then block kernel attempts to copy from/to opaque references. 
To do so, we associate the tagged kernel pages (i.e. contain opaque references) with pre-allocated user pages which only have dummy filedata, %
and direct all accesses to tagged pages to them.
With all changes reflected on tagged kernel pages and dummy user pages, OS is oblivious to opaque references and makes decision based on its intact logic (e.g. whether to flush dirty pages).
This transparently bypasses the page cache layer without disruptive changes.

As a result of the above modifications:
1) on the filedata write path, OS allocates kernel pages which preserve opaque references, and copies dummy user pages (i.e. filedata to write) to them. 
After \fs{} execution, OS generates bios whose filedata points to these kernel pages. 
It then submits the bios to TEE, returning opaque references. 
2) on the filedata read path, it is a mirror process.

\paragraph{Block translation tables}
A bio request received by the TEE carries a buffer address and a virtual block number. 
The latter is translated to a block number for the isolated physical disk.
The TEE does the translation by consulting with its per-image block translation tables (BTT). 
A BTT maps an OS-visible virtual block number to a TEE-visible physical block number.
BTT entries are only for metadata blocks. 
Filedata block numbers do not need translation -- 
they are either directly mapped to the physical disk or discarded.

BTTs reduce the cost of manipulating sybil \fs{} images. 
(1) Much of FIDS becomes BTT operations. 
To fork an image, the TEE duplicates its BTT without duplicating the disk blocks. 
To shuffle two images, the TEE unmounts the images, shuffle their BTTs, and re-mounts them. 
To retire an image, the TEE frees its BTT. 
(2) The TEE implements CoW by setting BTT entries of identical metadata blocks pointing to the same physical copy. 
When any shared disk block is written to, the TEE allocates a new disk block and updates BTT entries for all \fs{}s. 

\paragraph{Store BTTs securely}
The TEE stores encrypted BTTs in the normal world. There are two reasons: 1) BTTs should enjoy equal confidentiality and reliability as user files; 
2) storing BTTs on an in-TEE \fs{} (with crash consistency, etc.) would defeat our goal of leaving \fs{} out of TEE.

Outsourcing BTT storage leaks no secrets: 
BTT lookups are driven by disk requests submitted from the OS; 
the input block numbers are from the OS; 
the output block numbers will not be decrypted until they are in the TEE. %
The TEE updates BTTs only in an egalitarian fashion. 
In shuffling \fs{} images, the TEE re-encrypts all their BTT entries. %
To allocate a new disk block for a virtual block, the TEE re-encrypts BTT entries corresponding to the virtual block of \textit{all} \fs{}s. %

\paragraph{Metadata vs. filedata}
The following details are from real-world \fs{}s. 
(1) Because \fs{} logic needs to access directory contents, 
\sys{} treats directories as metadata, 
although they may be implemented as special files by some \fs{}s. 
(2) \sys{} treats a journaling \fs{}'s journal as metadata. 
By default, common journaling \fs{}s write dirty metadata (e.g. inodes) to their journals. 
As an expensive option, they can write filedata to journals for stronger consistency. 
In our current implementation we turn the option off. 
(3) Some OS functions may read filedata, e.g. exec() will parse the header of an executable file. 
These functions, however, are not supposed to be invoked on TEE-owned files (e.g. OS should not exec() a TEE program file). 
TEE can safely reject the read attempts.

%% file: sec-analysis.tex
\section{Security analysis}
\label{sec:eval:sec}

\input{tab-sec}

\subsection{TCB}
\sys{} keeps substantial OS code out of the TEE: 37K for \ext{}, 22K for \ffs{}, and 37K for a block layer, as reported by SLOCCount~\cite{sloccount} in kernel v4.19.
The \sys{} runtime only adds 1K SLOC to the TEE and its replay-based MMC driver adds another 1K SLOC. 
The TEE exports two interfaces to the OS, for issuing file calls and  for receiving disk requests.
Through the two interfaces, the normal/secure workloads share no state.
Following a common practice~\cite{ginseng},
the TEE passes messages with arguments packed into CPU registers during world switches, 
minimizing the risks of data leak. 
The only input data to the TEE is metadata. 
The TEE is secure against invalid or malformed metadata:
1) the TEE simply moves the metadata between OS buffers and the physical disk;
it never touches the metadata;  %
2) the TEE sets the backing memory to be non-executable. 

\subsection{Security guarantees}
\label{sec:security-analysis}

\paragraph{Against random guess attacks}
In each attempt, the attackers randomly pick one of $K$ \fs{} images and infer secret based on the file calls on the image.
They break the obfuscation if they either hit the actual image, or hit a sybil image that coincides with the actual secret.
Hence the probability to break \sys{} is $\frac{1}{K} + \frac{K-1}{K}(1/N)$, where $N$ is the number of plausible secrets represented by the trace library. As developers generate traces towards an $N$ much higher than $K$ (\S\ref{sec:trace:case}), e.g. $N$=500 secrets for \textit{Database},
the above probability is close to $1/K$.

Section~\ref{sec:eval} will compare \sys{} to ORAM when they provide same probabilistic guarantees.

\input{fig-fsnr-measure}

\paragraph{Against observation of file call histories}
We measure $P$: the probability that the OS pinpoints the history of a given \fs{} image F. 
$P$ is reciprocal to the number of images ($M$) that have histories indistinguishable from F to the view of the OS, due to shuffling (\S~\ref{sec:fsnr}). 
We model $M$ as a set cardinality: 
$M=|\bigcup_{i=1}^{r}S_i|$, where $r$ is rounds of shuffling, $S_i$ is the set of \fs{}s participating in $i$-th shuffling round;
$M \in [1,K] $.

We study how $P$ changes as a function of shuffling round $r$. 
In the above form of $P$, $S_i$ depends on the set of \fs{} images having identical metadata.
Therefore, we run all benchmarks with varying $K$, collect their traces, and report the average $P$ as a function of $r$. 
The results are shown in Figure~\ref{fig:fsnr-measure}. 
In general, as an image participates more rounds of shuffling, $P$ drops from 1 (OS is certain about its history) to lower values (less certain). 
$P$ drops faster with a larger number of images (higher K), 
because of richer shuffling opportunities and thus faster increase in uncertainty.

\paragraph{Against disk timing side channels}
We quantify the mitigation (\S\ref{sec:disk:timing}).
We measure the delays during 100 runs of \fs{} image initialization, consisting of over 400K physical disk accesses as described. 
We set the padding threshold to be 1.6 us, 99-percentile of the measured delays.

We then run a \fs{} with two versions: (1) \textit{Insecure}; (2) \sys{} with various paddings.
We use \textit{mutual information} (MI), a common notion of information theory used in privacy preservation~\cite{time-protection,trace-wringing}, to characterize the correlation between two random variables: (a) the access timing distribution of 1000 randomly sampled blocks on version 1 (i.e. \textit{Insecure}) and (b) the access timing distribution of the same set of blocks on version 2 (i.e. ours with various paddings).
Higher MI implies the higher certainty that the second timings are sampled from an emulated disk. %

Our measurement shows a significant MI reduction, from 0.342 bits (i.e. \sys{} w/o padding) to 0.007 bits.
Such a level of residue MI is considered negligible in prior work~\cite{time-protection}, 
suggesting it is difficult for the OS to correlate whether a disk is emulated to the observed access delays.

\paragraph{Against other attacks}
Table~\ref{tab:sec} shows that \sys{} leaks no filedata or activities to an adversarial OS even when the OS deviates from the \fs{} logic or injects malformed metadata. %
\sys{}'s confidentiality is susceptible to hardware side channels, e.g. through CPU cache. 
We rely on existing mitigations~\cite{time-protection}. 
\sys{} defeats filedata integrity attacks by rejecting OS accesses (\S~\ref{sec:disk:reject}).
\sys{} can detect availability attacks but cannot prevent them, e.g. the OS powering down the whole device.

%% file: tab-sec.tex
\begin{table*}[!h]
	\footnotesize
	\centering
	\begin{tabular}{l|l|l|c|c|c} 
	& Attacks                                   & Defense/mitigation                                  & C & I & A  \\ 
	\hline
	\multirow{4}{*}{\rotatebox[origin=c]{90}{\parbox[c]{1cm}{\centering Passive}}} & Observing file activities                 & Sybil filesystems with credible activities \S\ref{sec:disk} \& \S\ref{sec:sybil}                  	& \checkmark 	& - 	& -  \\
	\cline{2-6}
	& Observing filedata move                   & Isolating filedata path \S\ref{sec:disk:isolate}                                    	& \checkmark 	& -  	& -   \\ 
	\cline{2-6}
	& Measure disk timing                       & Delay padding \S\ref{sec:disk:timing}                                              	& \checkmark		& - 	& -  \\
	\cline{2-6}
	& Learning history of file activities       & Filesystem identity shuffling \S\ref{sec:fsnr}                              	& \checkmark 	& - 	& -  \\ 
	\hline
	\multirow{7}{*}{\rotatebox[origin=c]{90}{\parbox[c]{1cm}{\centering Active}}}  & Dropping file calls                       & TEE checks file integrity~\cite{trustshadow} 	& - 	& $\circ$ & $\circ$  \\ 
	\cline{2-6}
	& Dropping (un)mounting requests            & TEE checks filesystem integrity~\cite{sprobes}                      	& - 	& $\circ$ & $\circ$  \\ 
	\cline{2-6}
	& Unsolicited reads from filedata           & Reject OS access \S\ref{sec:disk:reject}                                           	& \checkmark		& - 	& -  \\ 
	\cline{2-6}
	& Unsolicited writes to filedata            & TEE erases existing filedata \S\ref{sec:disk:reject}; detect by file check~\cite{trustshadow}                    	& \checkmark 	& $\circ$ & -  \\ 
	\cline{2-6}
	& Unsolicited writes to metadata            & TEE not touching metedata \S\ref{sec:disk:isolate}; detect by filesystem check~\cite{sprobes}	& \checkmark 	& $\circ$ & -  \\ 
	\cline{2-6}
	& Supplying wrong block numbers of filedata & TEE checks file integrity~\cite{trustshadow}                       & -  	& $\circ$  &  -  \\
	\cline{2-6}
	& Fabricated references of filedata         & Strong opaque references \S\ref{sec:disk:path}              & \checkmark 	& \checkmark & -  \\
	\end{tabular}
	\caption{\sys{} thwarts attacks against confidentiality. \textbf{C:} Confidentiality, \textbf{I:} Integrity, \textbf{A:} Availability. \textbf{\checkmark:} Attack thwarted, \textbf{$\circ$:} Attack detected \textbf{-:} Not targeted by the attack}
	\label{tab:sec}
	\vspace{-2em}
\end{table*}

%% file: fig-fsnr-measure.tex
\begin{figure}[h]
	\centering
	\begin{minipage}{0.48\textwidth{}}
		\includegraphics[width=\textwidth{}]{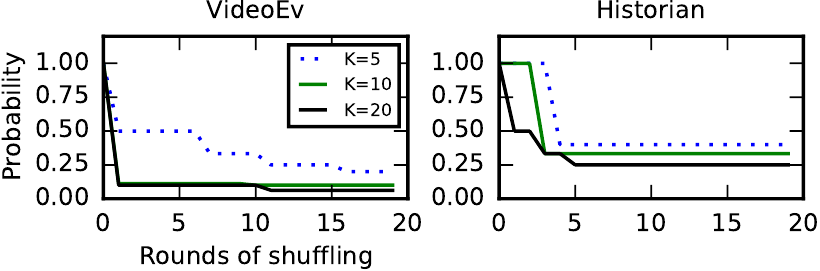}
		\vspace{-2em}
		\caption{FIDS diminishes probability of OS distinguishing an image's history from histories of other images. Write-most benchmarks are shown. On read-most benchmarks the probability converges even faster}
		\label{fig:fsnr-measure}
	\end{minipage}
	\vspace{-1em}
\end{figure}

%% file: exp.tex
\vspace{-3mm}
\section{Evaluation}
\label{sec:eval}
We seek to answer the following questions on \sys{}: 

\begin{myitemize}

\item How much is the space overhead?
(\S\ref{sec:eval:space})

\item How much is the slowdown in file accesses?
(\S\ref{sec:eval:delay})

\item 
What is the performance impact of FIDS?
(\S\ref{sec:eval:fsnr})

\end{myitemize}

\input{bench}

%% file: bench.tex
\subsection{Methodology}
\label{sec:eval:bench}

\input{exp-setup}

\input{fig-emu-storage}

\input{strawman}

\input{macrobenchmarks}

%% file: exp-setup.tex
\input{tab-strawman2}

\input{fig-exp-setup}
\paragraph{Setup and metrics}
Table~\ref{tab:plat} summarizes our test platform.
We choose Rpi3~\cite{rpi3} for its good support for~\tz{}. 
Table~\ref{tab:benchmarks} summarizes the trustlets as benchmarks and their traces.
In TrustZone, we do not run them but extract their file traces for replay, making benchmarks simple and reproducible. 
Note that our benchmark programs are for reproducing trustlets' file activities; 
production trustlets likely have different, more compact implementations, e.g. by linking to embedded libraries.
We deliberately diversify file behaviors and file sizes. 
For each benchmark, we create the smallest disk partition that can accommodate the benchmark files. %
Note that the smallest disk sizes supported by \ffs{} and \ext{} are 39 MB and 2 MB. %

\paragraph{\Fs{} choices}
We pick two mainstream \fs{}s that exercise \sys{} in different ways. 

\begin{myitemize}
\item 
\ffs{} is \sys{}'s reference \fs{}. 
A log-structured \fs{} popular on mobile devices, F2FS extensively optimizes for NAND flash~\cite{f2fs}. 
For flash longevity, \ffs{} allocates blocks on demand and generates compact metadata. %
We create the test image with mkfs.f2fs v1.11.0.

\item 
\ext{} is our stress test for \sys{}. 
A journaling \fs{}, \ext{} issues dense metadata writes 
\cite{union-buffer-cache-fast}. 
Since \sys{} stores metadata for sybil images, 
it incurs higher overhead with \ext{}. 
We create the test image with mkfs.ext4 v1.44.5.

\end{myitemize}

%% file: tab-strawman2.tex
\begin{table*}[h]
	\vspace{1em}
	\footnotesize
	\centering
	\begin{tabular}{l|l|l|l|l|l}
		\multicolumn{1}{c|}{\textbf{Trustlets \& description}}                                                                                 & \multicolumn{1}{c|}{\textbf{IO Char.}}                        & \multicolumn{1}{c|}{\textbf{Dataset }}                                                                                                                             & \begin{tabular}[c]{@{}l@{}}\textbf{Access}\\\textbf{delay }\end{tabular} & \multicolumn{1}{l|}{\begin{tabular}[c]{@{}l@{}}\textbf{Side}\\\textbf{channels}\end{tabular}} & \multicolumn{1}{l} {\begin{tabular}[c]{@{}l@{}}\textbf{Comparisons}\\\textbf{(Baselines)}\end{tabular}}                  \\
		\hline
		\begin{tabular}[c]{@{}l@{}}\textbf{Database.} Query on-device\\database of user health data.\end{tabular}         & \begin{tabular}[c]{@{}l@{}}Single file\\ Rand RW\end{tabular} & \begin{tabular}[c]{@{}l@{}}select(1-8) benchmarks from SQLite~\cite{sqlite-speedtest}. \\Total: 500 queries, 14K file calls. File size: 800KB\end{tabular}                & \begin{tabular}[c]{@{}l@{}}Per\\ query\end{tabular}                         & Sizes \& offsets                                                                                              & \multirow{3}{*}{ORAM}                                        \\
		\cline{1-5}
		\begin{tabular}[c]{@{}l@{}}\textbf{Fulltext.} Search text files for \\ on-device QA.\end{tabular}                       & \begin{tabular}[c]{@{}l@{}}Multi files\\ Rand RD\end{tabular} & \begin{tabular}[c]{@{}l@{}}Lucy~\cite{apache-lucy} on pre-indexed Enron emails~\cite{enron}.\\ Total: 100 queries; 100K file calls. File size: 2 GB\end{tabular} & \begin{tabular}[c]{@{}l@{}}Per\\ search\end{tabular}                        & Sizes \& offsets                                                                                             &                                                              \\
		\cline{1-5}
		\begin{tabular}[c]{@{}l@{}}\textbf{ModelLoad.} Load ML models \\ from files.\end{tabular}                                 & \begin{tabular}[c]{@{}l@{}}Multi files\\ Rand RD\end{tabular} & \begin{tabular}[c]{@{}l@{}}TensorflowLite~\cite{tensorflow-web} loading 10 neural nets, \\Total: 80K file calls. File size: 41 MB.\end{tabular}                         & \begin{tabular}[c]{@{}l@{}}Load per\\ NN\end{tabular}                       & \begin{tabular}[c]{@{}l@{}}Sizes \& offsets\\ File paths \end{tabular}                                                                                           &                                                              \\
		\hline
		\begin{tabular}[c]{@{}l@{}}\textbf{Historian.~}Log data bags \\from multi. sensors on a robot.\end{tabular}              & \begin{tabular}[c]{@{}l@{}}Single file\\ Append\end{tabular}  & \begin{tabular}[c]{@{}l@{}}ROSBag on EU Long-term dataset~\cite{rosbag}.\\Total: 36K file calls. File size: 659 MB.\end{tabular}                                & \begin{tabular}[c]{@{}l@{}}Log per\\ data bag\end{tabular}                  & Sizes \& offsets                                                                                             & \begin{tabular}[c]{@{}c@{}}PadWrite\end{tabular}      \\
		\hline
		\begin{tabular}[c]{@{}l@{}}\textbf{VideoEv.} Log images of \\motion events detected.\end{tabular}  & \begin{tabular}[c]{@{}l@{}}Multi files\\ Seq WR\end{tabular}  & \begin{tabular}[c]{@{}l@{}}50 1080P images in Bangor video~\cite{bangor-webcam}.\\Total: 9K file calls. File size: 100 KB\end{tabular}      & \begin{tabular}[c]{@{}l@{}}Log per\\ image\end{tabular}                     & Access occurrences                                                                                            & \begin{tabular}[c]{@{}c@{}}InjectCreate\end{tabular}  \\
		\hline
		\begin{tabular}[c]{@{}l@{}}\textbf{CredLoader.} Load credentials \\ for authentication with servers.\end{tabular} & \begin{tabular}[c]{@{}l@{}}Multi file\\ Seq RD\end{tabular}   & \begin{tabular}[c]{@{}l@{}}Load 50 key files generated with ssh-keygen.\\ Total: 150 file calls. File size: 0.9 KB.\end{tabular}                                   & \begin{tabular}[c]{@{}l@{}}Load per \\ key\end{tabular}                     & File paths                                                                                               & \begin{tabular}[c]{@{}c@{}}InjectFiles\end{tabular}   \\
	\end{tabular}
	\caption{A summary of benchmarks}
	\label{tab:benchmarks}
	\vspace{-2em}
\end{table*}

%% file: fig-exp-setup.tex
\begin{table}
\centering
	\includegraphics[width=0.48\textwidth{}]{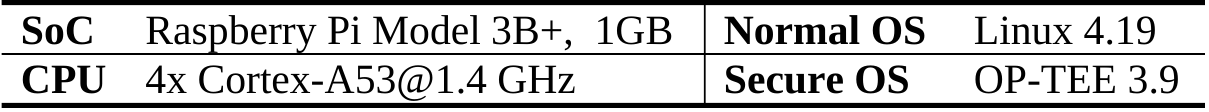}
	\caption{The test platform used in evaluation}
	\label{tab:plat}
	\vspace*{-2em}		%
\end{table}

%% file: fig-emu-storage.tex
\begin{figure*}[t!]
	\vspace*{0.5em}
	\centering
	\includegraphics[width=\textwidth{}]{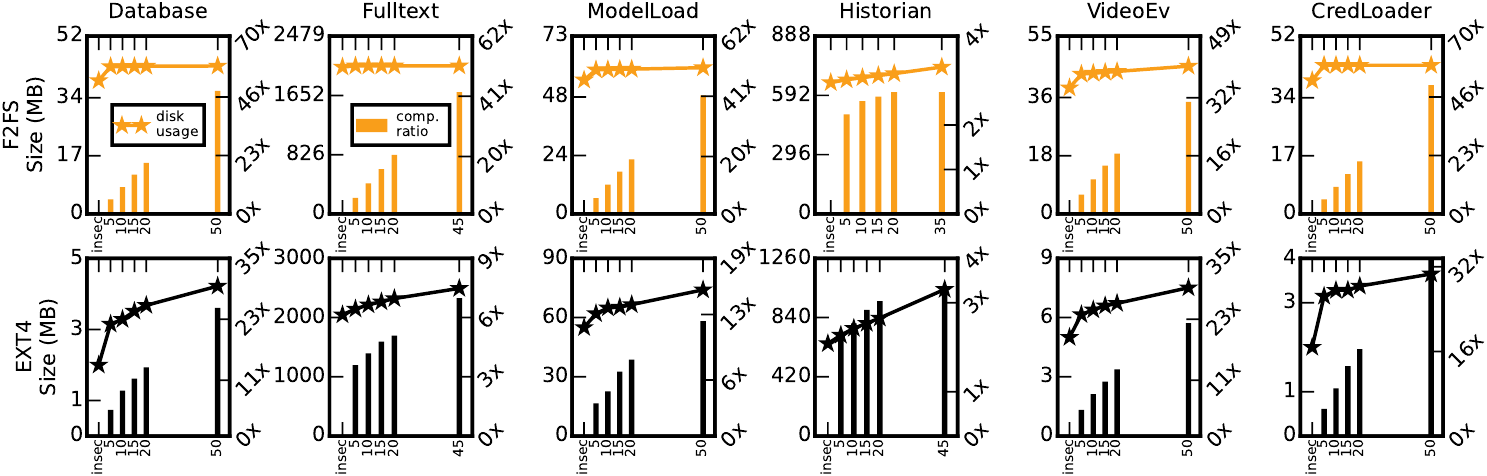}
	\vspace{-2em}
	\caption{Disk usage (lines) and metadata compression ratio (columns). \sys{}'s usage is modestly higher than \textit{Insecure} and grows gracefully with K on most benchmarks.  X axis: number of \fs{} images (K).  Note that an image of F2FS/Ext4 has a minimum size of 2 MB/39MB by design
	}
	\label{fig:emu-storage}
	\vspace{-2mm}
\end{figure*}

%% file: strawman.tex
\paragraph{Baselines}
First, we consider \textit{Insecure} which incurs no overhead: 
the TEE invokes a \fs{} image without any protection. %
Furthermore, 
we consider protection baselines that specifically hide the side channel of each trustlet. 
As such, these protections pay no cost for unneeded protection and are therefore competitive against \sys{}. 
They are summarized in Table~\ref{tab:benchmarks} with details as follows.

	\begin{myitemize}
	\item 
	\textit{ORAM}~\cite{obliviate} %
	is the baseline protection for Database, FullText, and ModelLoad. 
	It mitigates side channels due to access sizes and offsets within a single file through obfuscation.
	By design, ORAM guarantees that the probability of random guesses recovering the file trace is
	$P=1/2^{LM}$, 
	 where $L$ is the height of ORAM tree and $M$ is number of accesses in the trace~\cite{path-oram}. 
	 Since \sys{} provides $P=1/K$,  
	we compare the overheads of ORAM and \sys{} when their guarantees match, i.e. $P=1/2^{LM}=1/K$. 
	We set $K=50$, ($P=0.02$), the largest number of \fs{} images \sys{} can run on our test platform with limited TEE memory. 
	Since ORAM's $L$ and $M$ must be integers, we choose the closest value $LM=6$, where $M$ depends on a benchmark's trace segments, e.g. $M=3$ for \textit{Database}.

	\item 
	\textit{PadWrite} hides the append sizes for \textit{Historian}. 
	The TEE pads the size of each append to be the largest append the benchmark may issue. 
	No sybil files or calls are injected. 
			
	\item 
	\textit{InjectCreate} hides file creation occurrences for \textit{VideoEv}. 
	The TEE creates the same number of sybil files as the actual files. The creation times are independent of the actual creation. 
	Since \textit{VideoEv}'s file sizes are not secret, the TEE creates the sybil files with same sizes as the actual.
	
	\item 
	\textit{InjectFiles} hides file paths for \textit{CredLoader}. 
	The TEE injects the same number of sybil files as the actual files in the actual image and emits sybil reads to them. 
	Since \textit{CredLoader}'s access offsets within files are no secret, the sybil reads use the actual offsets. 
					
	\end{myitemize}

%% file: macrobenchmarks.tex
\vspace{-2mm}
\subsection{Space overhead}
\label{sec:eval:space}

\paragraph{Disk overhead}
comes from (1) the metadata size per sybil image amplified by (2) the number of sybil images ($K-1$).

As shown in Figure~\ref{fig:emu-storage}, the disk overhead is modest in most benchmarks.
Compared to \textit{Insecure}, \sys{} increases the disk usage by 1\%-58\% (18\% on average) when $K=5$.
When $K$ reaches as high as 50,
the disk space of \sys{} as compared to \textit{Insecure} is 38\% on average,
which roughly translates to 1\% per additional sybil image.

Our experiments show the efficacy of the metadata CoW compression ~(\S\ref{sec:disk}).
For example, with \K{}=20, turning off CoW increases the disk overhead by 2$\times$ on \ffs{} and 4$\times$ on \ext{}.
When we further turn off discard of filedata, the disk overhead is almost linear to $K$. %

\input{tab-strawman-storage}

\paragraph{Comparison with baselines}
Table~\ref{tab:strawman:storage} shows their minimal disk usage.
\textit{ORAM} incurs 9$\times$ disk overhead, 3$\times$ to 9$\times$ higher than \sys{} with $K=50$,
which is consistent with prior ORAM-based file protection~\cite{obliviate}.
This is because ORAM-based protection must store the whole ORAM tree, several times larger than the address space (file size) to be protected.

For trustlets that can be protected with simple obfuscation,
the baselines may use less disk space than \sys{}. 
For instance, on \textit{Historian} and \textit{VideoEv} which append to a single file, 
their disk usage is 6\% and 18\% lower than \sys{} with $K=20$. 
This is because 1) intensive filedata writes update metadata frequently, which makes \sys{}'s compression less effective;
2) the metadata on many sybil images exceeds the total size of small files (e.g. 5MB).

\paragraph{Memory overhead} of \sys{} is from storing BTTs and the metadata of sybil images. 
Such memory consumption grows with $K$. 
It is allocated in the normal world only and the stable consumption is modest,
e.g. 26 MB and 18 MB for running \textit{ModelLoad} on \ext{} and \ffs{} when $K=20$, a small fraction of the 1GB DRAM on our board.

\vspace{-2mm}
\subsection{File access delays}
\label{sec:eval:delay}

We measure delays of file access sequences as defined in Table~\ref{tab:benchmarks}.
The rationale is the mobile/embedded trustlets are often event-driven and latency-sensitive.

\input{fig-e2e-macro}

Figure~\ref{fig:e2e-macro} shows the results.
On most benchmarks, the delays grow gracefully with $K$.
Compared to \textit{Insecure} with only the actual image:
\sys{} with $K=20$ increases the delays by 1.3$\times$-4.5$\times$ (2$\times$ on average), showing a sublinear growth. 
\sys{} benefits from its elimination of filedata access for sybil images, 
which discards 95\% of disk requests on average. %
The delay of \textit{VideoEv} does not grow because 
the TEE issues file calls to images at different times in order to hide access occurrence; these file calls do not contend. 
When $K>20$, \sys{}'s concurrent execution of \fs{}es is bound by four ARM cores on our board.
For \textit{Historian} and \textit{Fulltext} on \ffs{} with $K$ > 45, our test board runs out of TEE memory. %
Note that the delays of normal/secure world switches (i.e. ns) are negligible compared to the disk IO delays (i.e. us).

\paragraph{Comparisons}
Compared to ORAM, \sys{}'s delays are lower by
8$\times$ -- 70$\times$ (on average 37$\times$). %
In contrast to \sys{} which \textit{discards} sybil filedata, \textit{ORAM} \textit{amplifies} filedata by tens of times. 
While a benchmark reads tens of MBs of data, \textit{ORAM} amplifies reads by 18$\times$ and adds 20$\times$ extra writes. 
This results in 40$\times$ disk IO and saturates our disk bandwidth.

On \textit{Historian}, \sys{}'s delay up to 4$\times$ higher than \textit{PadWrite}. 
This is expected, as \textit{PadWrite} 
precisely caters to \textit{Historian}'s append-only pattern and its side channel. %
Its only overhead is write of additional filedata. 
On \textit{CredLoader} with F2FS, the delay with $K=50$ is 63\% higher than the baseline \textit{InjectFiles}. 
The reason is in F2FS's low performance in accessing many small files of this benchmark.

\input{bench-fsnr}

%% file: tab-strawman-storage.tex
\begin{table}[h]
	\centering
	\includegraphics[page=1,width=0.48\textwidth{}]{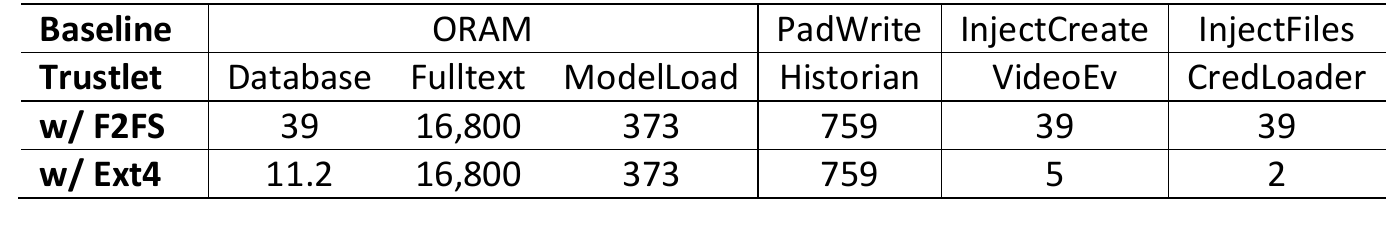}
	\caption{Disk space (MB) needed by baselines. 
	Note that EXT4/F2FS have the least allowable disk sizes of 2MB/39MB respectively.
	Figure~\ref{fig:emu-storage} shows the disk usage of \sys{}
	}
	\label{tab:strawman:storage}
	\vspace{-2em}
\end{table}

%% file: fig-e2e-macro.tex
\begin{figure}[h]

	\centering
	\includegraphics[width=0.48\textwidth{}]{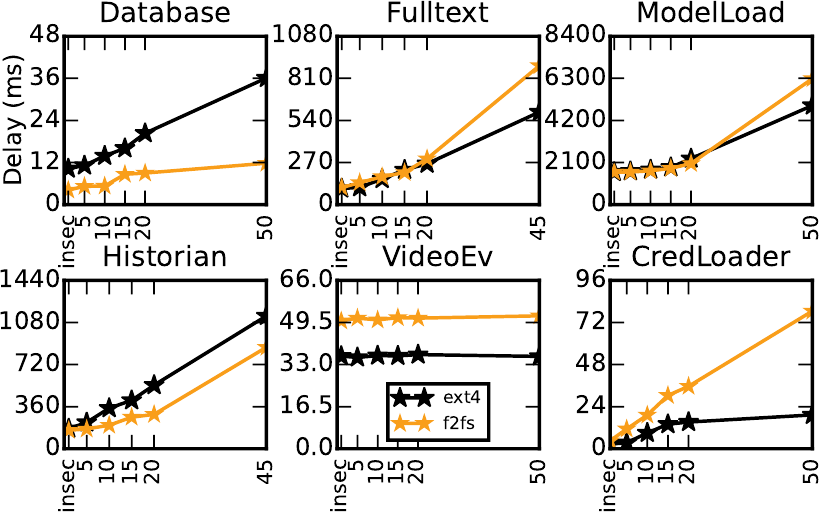}
	\vspace{-2em}
	\caption{File access delays. 	\sys{}'s delays are modestly higher than \textit{Insecure} and grow gracefully with K on most benchmarks.
	X axis: number of \fs{} images (K).  Delay metrics defined in Table~\ref{tab:benchmarks}}
	\label{fig:e2e-macro}

	\includegraphics[page=1,width=0.48\textwidth{}]{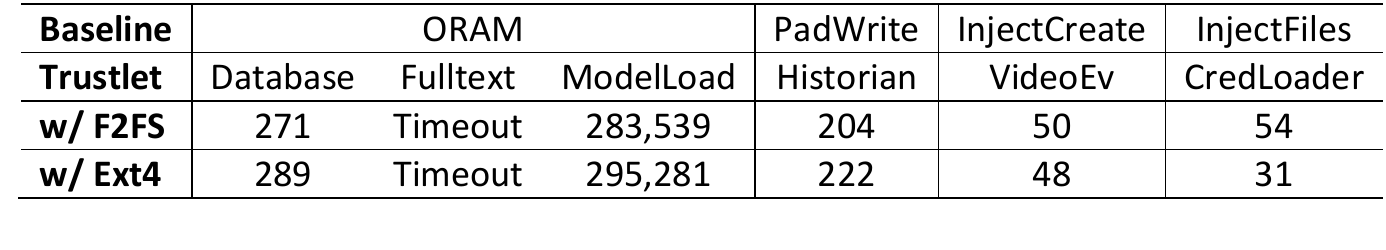}
	\vspace{-2em}
	\captionof{table}{File access delays (in ms) by baselines for comparison. \sys{}'s results are in Figure~\ref{fig:e2e-macro}}	
	\label{tab:strawman:delay}
	\vspace{-2em}
\end{figure}

%% file: bench-fsnr.tex
\input{fig-fsnr-perf}

\subsection{FIDS overhead}
\label{sec:eval:fsnr}

\paragraph{Costs of FIDS operations}
Shuffling and forking do not require data copy or move. 
Of their delays, 10\%--30\% comes from BTT manipulation
while the remaining comes from stopping and restarting \fs{} images as done by the untrusted Linux kernel. 
With \ext{}:
1) forking takes 90ms/180ms on a 64MB/2GB disk, respectively. 
2) shuffling two disks of 64MB/2GB each takes 80ms and 130ms, respectively.
3) retiring an image takes less than 1 ms.
Compared to \ext{}, 
F2FS shows 39\% -- 66\% shorter delays due to its ``fastboot'' option. 
Our measurements suggest FIDS efficiency can benefit from further optimization of the Linux kernel, e.g. by parallelizing mounting/unmounting of many \fs{} images.

\paragraph{Impact on trustlet throughputs}
Because by design FIDS is executed in the background off the file access path, 
we focus on its impacts on a trustlet's throughputs.

Figure~\ref{fig:fsnr-perf} shows three benchmarks where throughput matters. 
We chose the intervals based on~\cite{android-io}, which reports low-frequency and bursty file activities for mobile/edge device;
under such intervals, at most 1-2 secrets may be exposed even in case of accidental \fs{} identity leak.
We validate their throughputs are bound by disk IO because the throughputs are higher when running them on \textit{Insecure} \fs{}. 
Hence, our test trustlets do not execute app logic, e.g. database code; 
they execute file accesses as quickly as \sys{} allows.

Even under strong protection (e.g. $K=20$; FIDS every second),
the benchmarks deliver throughputs appropriate to the IoT/embedded scenarios.
\textit{Database} and \textit{Fulltext} can process tens of queries per second and 
several queries per second, respectively, sufficient to queries driven by a single user. 
\textit{Historian} can log a few MBs of data per second, which can support a robot's 1--2 HD video streams or point clouds at 3--5 FPS~\cite{chinchali2019network}. 
As the developers relax the protection, 
the throughputs improve by 1.5$\times$ to 2$\times$.
Using \textit{Insecure} increases these throughputs by 2$\times$-10$\times$.

%% file: fig-fsnr-perf.tex
\begin{figure}[h]
	\includegraphics[width=0.48\textwidth{}]{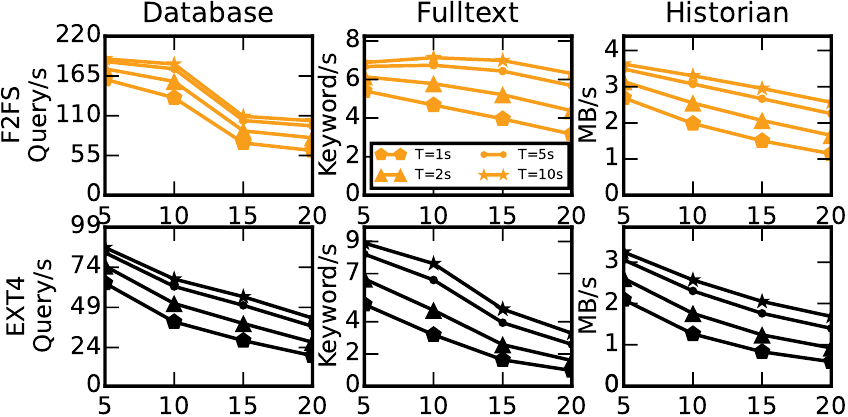}
	\vspace{-2em}
	\caption{Trustlet throughputs under different FIDS intervals (T). X axis: number of images (K)
	}
	\label{fig:fsnr-perf}
\end{figure}

%% file: related-work.tex
\section{Related Work}
\label{sec:related}

\paragraph{Side-channels \& mitigations}
Timing side channels are often mitigated by deploying low-res timer~\cite{vattikonda_eliminating_2011,mozilla_clamp_2015}, padding delays~\cite{time-protection}.
We do not focus on them but apply these techniques to mitigate the side channel of our emulated disk.
Access pattern side channels (e.g. memory~\cite{controlledChannel,access-patten-privacy}, file~\cite{obliviate}) exploit data-dependent execution to infer user input (e.g. queries).
They are often mitigated by distorting the access pattern (e.g. via ORAM~\cite{oram,opaque,oblidb}).
Motivated by them, we also protect access pattern;
unlike them, we preserve the patterns yet hide them under credible sybil ones.

\paragraph{Hide data in plain sight}
To hide data that must eventually be released, an old wisdom is to add noise~\cite{noise-addition} as deception.
Due to its practicality, recent systems start to adopt it for anonymous location sharing~\cite{sybilquery}, query processing~\cite{kloakdb}.
Compared to them, we are the first to apply it to file services.
Another approach is to continuously reset attacker's observations on the data (e.g. ASLR~\cite{aslr} limits attacker's observation on address spaces). 
We echo its motivations;
we deal with filesystem identities (actual vs. sybil), a different domain.

\paragraph{TEE and file services} 
To enable files services for these apps, some include \fs{} code inside TEE (e.g. through porting~\cite{haven}, libraryOS~\cite{graphene-sgx}, build anew~\cite{besfs}).
Compared to them, we do not include nor invent \fs{} code, instead we take a forwarding approach which reuses unmodified \fs{} code.
Some forward file calls as we do~\cite{trustshadow, shinde_panoply_2017}.
In comparison, we focus on the ignored side channel caused by such forwarding.
Some exposes to apps a raw block device interface~\cite{sbd}, which is backed by a file in normal OS for crash consistency.
Similar to it, we also store a backing file in normal OS (i.e. BTTs).
Different from it, we store inside TEE a compact representation of \fs{}s, similar to David~\cite{david}.

%% file: conclusion.tex
\section{Concluding remarks}
\label{sec:conclusion}

\paragraph{Applicability to SGX}
\sys{} may hide the file activities of an SGX TEE (enclave) albeit with different implementation requirements.
Unlike a TrustZone TEE, an SGX enclave lacks the capability of direct disk access.
Thus, \sys{} may use a hypervisor to manage disk hardware for the enclave and 
isolate the disk from the untrusted OS, similar to~\cite{bitvisor,twindrivers}.

\paragraph{Conclusions}
\sys{} hides file activities of a TrustZone TEE. With \sys{}, the TEE generates sybil calls by replaying; the TEE backs only one image with the actual disk while other images with emulated storage; the TEE prevents the OS from learning long history of any image.
We build \sys{} and show that \sys{} works with unmodified file systems, incurs affordable overhead, and represents a new design point in guarding IoT storage stack. 
\sys{} opens the door for a TEE to external untrusted OS services.